\documentclass[
	letterpaper, % Paper size, use either a4paper or letterpaper
	10pt, % Default font size, can also use 11pt or 12pt, although this is not recommended
	twoside, % Two side traditional mode where headers and footers change between odd and even pages, comment this option to make them fixed
]{LTJournalArticle}

\runninghead{Metacognition is All You Need} % A shortened article title to appear in the running head, leave this command empty for no running head

\footertext{\textit{https://replicantlife.com}} % Text to appear in the footer, leave this command empty for no footer text

\usepackage{titling}

\title{Metacognition is all you need? \\\Large Using Introspection in Generative Agents to Improve Goal-directed Behavior } % Article title, use manual lines breaks (\\) to beautify the layout

\thanksmarkseries{arabic}

\author{%
  Jason Toy\thanks{Jason Toy, replicantlife.com - \href{mailto:jasontoy@gmail.com}{jasontoy@gmail.com}}, 
  Phil Tabor\thanks{Phil Tabor, neuralnet.ai - \href{mailto:phil@neuralnet.ai}{phil@neuralnet.ai}}, 
  Josh MacAdam
}

\renewcommand{\maketitlehookd}{%
	\begin{abstract}
		\noindent Recent advances in Large Language Models (LLMs) have shown impressive capabilities in various applications, yet LLMs face challenges such as limited context windows and difficulties in generalization. In this paper, we introduce a metacognition module for generative agents, enabling them to observe their own thought processes and actions. This metacognitive approach, designed to emulate System 1 and System 2 cognitive processes, allows agents to significantly enhance their performance by modifying their strategy. We tested the metacognition module on a variety of scenarios, including a situation where generative agents must survive a zombie apocalypse, and observe that our system outperform others, while agents adapt and improve their strategies to complete tasks over time. All code can be found at https://replicantlife.com
	\end{abstract}
}

\begin{document}

\maketitle % Output the title section

%----------------------------------------------------------------------------------------
%	ARTICLE CONTENTS
%----------------------------------------------------------------------------------------
\begin{figure}[]
\centering
\includegraphics[width=0.5\textwidth]{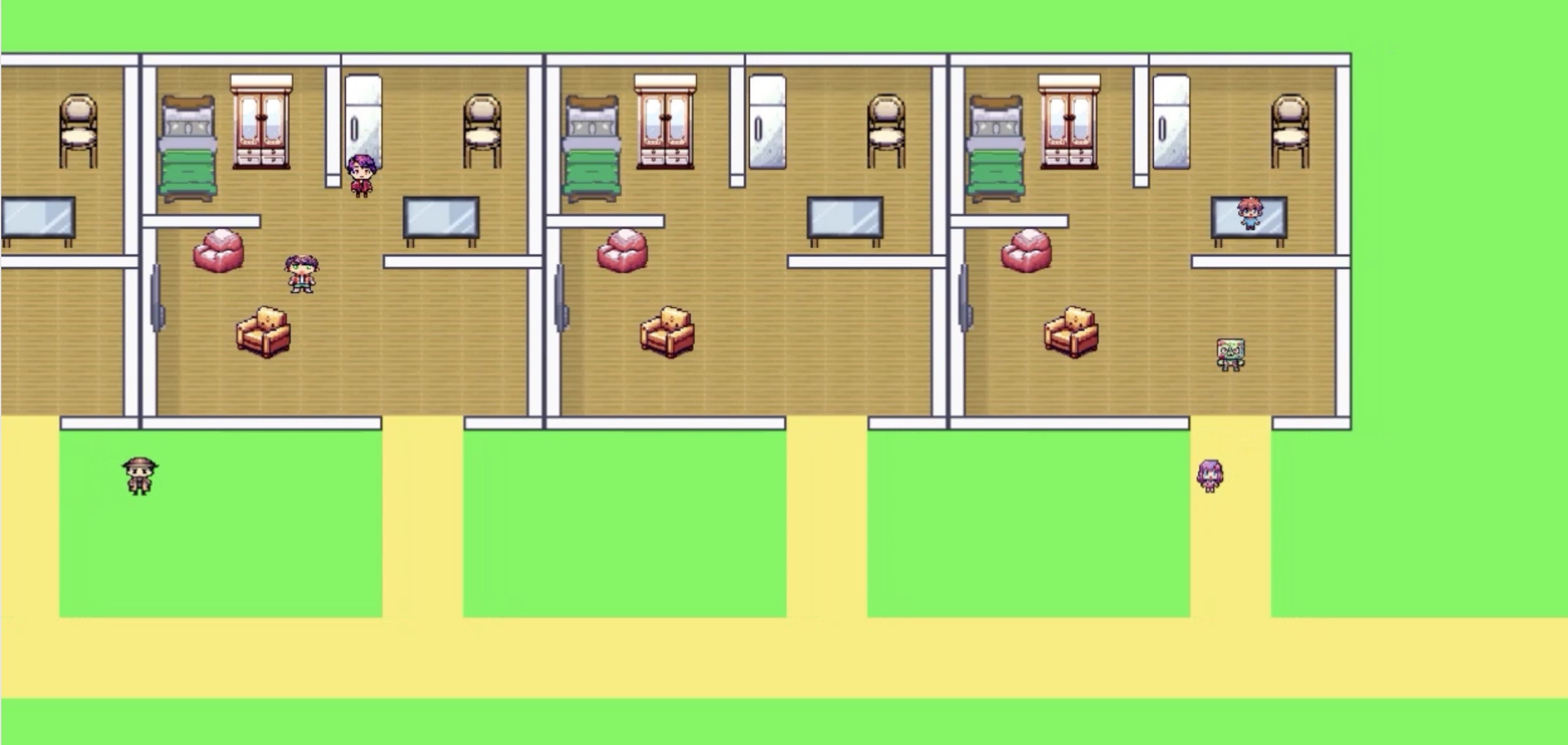}
\caption{Zombie Apocalypse Simulation}
\end{figure}

\begin{figure*}[ht]
    \centering
    \includegraphics[width=0.8\textwidth]{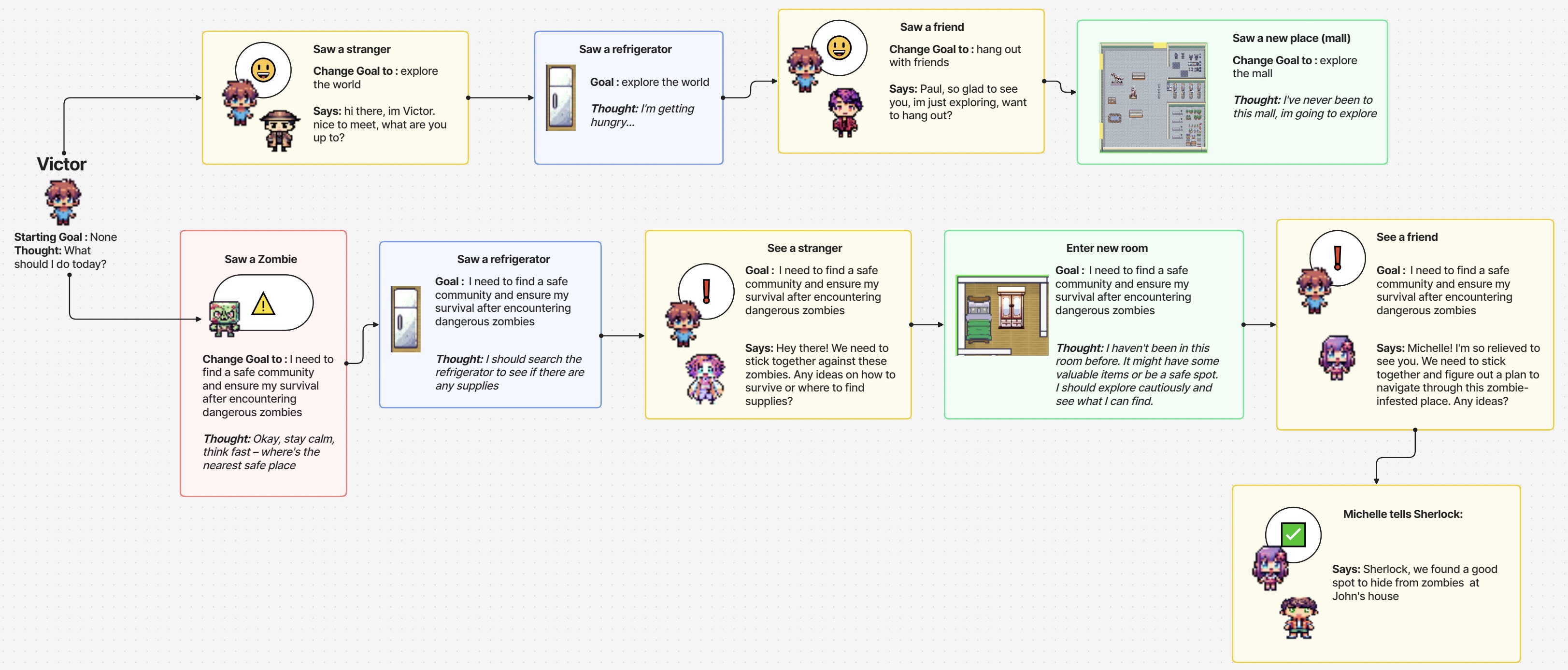}
    \caption{A timeline showing an agent's goals and thoughts change as it interacts in the simulation.}
    \end{figure*}

\section{Introduction}

Metacognition refers to the higher-order cognitive processes that involve thinking about one's own thinking. It encompasses a range of mental activities related to monitoring, regulating, and organizing cognitive processes to achieve specific goals. Metacognitive abilities enable individuals to reflect on their knowledge, problem-solving strategies, and learning experiences and therefore play a crucial role in shaping and modifying one's habits. For example, if one is studying for an exam, metacognitive processes might involve setting goals, choosing appropriate study strategies, monitoring their comprehension, and adjusting their approach if they are not understanding the material. 

The concept of System 1 and System 2 thinking, popularized by psychologist Daniel Kahneman, provides a framework for understanding metacognition\cite{Kahneman}. System 1 represents fast, automatic, and intuitive thinking, while system 2 involves slower, deliberate, and reflective thinking. In this framework, metacognition can be thought of as a specific System 2 process that examines actions from both System 1 and System 2 processing. Another analogy is system 1 is subconscious thinking and system 2 is conscious thinking (the voice you hear in your head).  Metacognition is a slow, expensive, and methodical thought process and is therefore better suited for introspective or strategic thinking, rather than immediate problem solving.

Metacognition in essence is "thinking about thinking" and requires the ability to look at one's own thoughts and thought processes from different points of view. Reflection, on the other hand, is typically characterized as looking at past experiences and deriving insights for future actions. Metacognition allows one to adjust their thought process and strategy based on asking relevant questions. Due to the dynamic nature of metacognition, no one strategy will be set in stone. Metacognition typically involves asking oneself different questions as a probing mechanism to further understand. Some questions one may see: "What do I know about this topic?", "Why do I want to achieve this goal?", "How can I monitor my progress towards my goal?", "How can I adjust my strategy to overcome current challenges?", "Metacognition is all you need?" etc. Metacognition is often applied to different types of thinking such as: problem solving, goal setting, reflection, learning, monitoring and evaluation, emotional regulation and other types of cognitive processes. As artificial agents are usually given a specific task, we focus on metacognition related to problem solving, monitoring, and evaluating progress towards their specific task.

Recent work with large language models has incorporated human cognitive processes into simulations of interacting generative agents tasked with cooperating to achieve strategic objectives \cite{qian2023communicative}. In particular, planning, memory, and reflection have been implemented in an effort to elicit human-like behaviors such as long term planning and cooperation among agents \cite{park2023generative}. The success of this work raises the question of what role metacognition may play in further enhancing the believability of behaviors of generative agents.

%------------------------------------------------

\section{Large Language Models with Cognitive Modules as Generative Agents}

Multiple experiments have incorporated metacognition into computational frameworks. \textcite{cox2022computational} outlines a general computational architecture in lisp. 
\textcite{mustafa2021assured} provides a framework for autonomous vehicles that adds a metacognition layer to monitor safety violations on top of generic reward accumulation.

\textcite{krueger2022enhancing} created a generic deep reinforcement learning (DRL) framework that incorporates metacognition into traditional reinforcement learning frameworks.
  
\textcite{park2023generative} demonstrated that LLMs equipped with reflection, observation, and planning modules on agents can successfully mimic believable human behavior in a simulated town environment. Seeding the idea of an agent wanting to host a valentine's party resulted in the agent successfully organizing a party which other agents discussed and attended.

We propose a metacognize module which allows agents to broadly contemplate their circumstances in order to create alternative strategies and improve performance. This subsumes the more tactical reflect and plan functions as presented in \textcite{park2023generative}. We show through ablation that agents can learn to continually adapt their strategies depending on the situation. Their overall strategy for dealing with problems affects the specific actions they take.

\begin{figure*}[ht]
\centering
\includegraphics[width=0.7\textwidth]{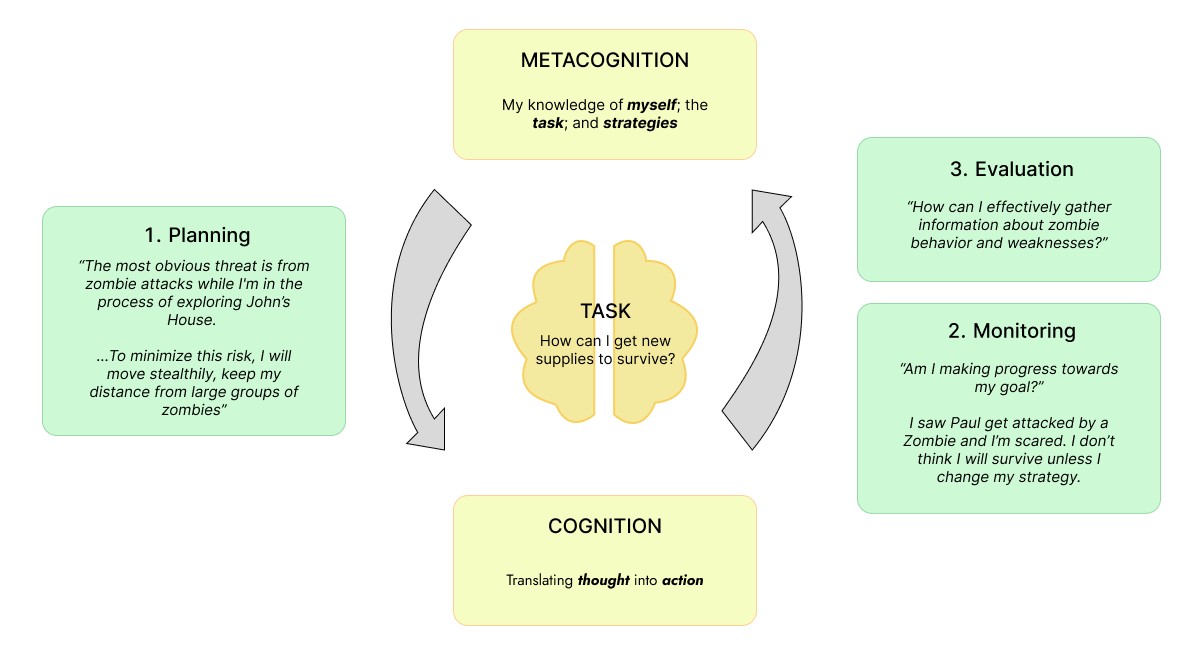}
\caption{Graphical representation of metacognition process.}
\end{figure*}

\section{Architecture}

We implement many of the same modules from \textcite{park2023generative}, with the addition of a group of modules dubbed \texttt{meta\_cognize}. As an agent progresses through the simulation, it accumulates a history of observations, memories, and thoughts. Agents are given goals but can optionally be left blank. When an agent starts towards its goal, it is not given an explicit strategy to follow. Instead, each agent periodically evaluates how it is progressing towards its goals by reviewing memories, thoughts, and past actions. The agent assigns itself a numeric score as well as a text statement for its reasoning for providing that score. This evaluation is stored in its memory as a meta-thought.

If the agent finds that it is not making enough progress, the agent calls its \texttt{meta\_cognize} module. When metacognition occurs, the agent asks itself how it might improve its performace in light of what it has learned. Additionally, the agent will periodically self-generate new introspective questions to think about its goals from different perspectives. For example, in the zombie apocalypse scenario, an agent initially starts with no goal or strategy, but after some time, we observe an agent contemplate these thoughts:

\begin{quote}
    \textit{``How can I survive this zombie apocalypse? What resources do I need? Where should I go for safety? How can I learn from both successes and failures to improve survival strategies? "}
\end{quote}

Depending on the current task and goals of the agent, those questions will change over time, influencing how the agent responds and acts in the environment. Agents have memory that is stored outside of the LLM, where each memory stores content, timestamp, location, importance score, and type of memory. Each time an agent reviews memories for higher cogntive function, memories are ranked by relevance to the speficic question it is considering. Relevance is calculated by cosine similarity of the question and memory embeddings.

Agents have two kinds of memories, a short term and long term memory. Short term memory stores a maximum of seven recent memories and is forgotten after approximately 30 seconds, modeled after human short term memory\textcite{1956miller}.  Long term storage memory is essentially unlimited and stored in system RAM.  Due to limited context window sizes, agents cannot process every memory when making decisions. This is similar to human memory where we may store large amounts of data, but only certain memories can be recalled at a time.

Retrieval-Augmented Generation (RAG) \textcite{lewis2021retrievalaugmented} is a technique created to give LLMs the ability to surface and reference content that the model was not explicitly trained on. 
Our memory recall system can be seen as a dynamic RAG system. As the agent's system acquires new memories and metacognizes over time, the relevant memories sent to the LLM change.

Anytime an agent makes a decision, relevant memories are retrieved to prime the agent on what to do.
There are several memory types such as observance memories like ``John saw a cat'' and conversation memories like ``John said 'How are you doing Paul?''. We explicitly store metacognition memories where an agent looks at its past memories and actions and asks a meta-question. That thought is stored as a meta-memory and inserted into the memory stream of the agent. In future actions and conversations, these meta-memories are recalled along with other memories to prime the agent to think about these meta-thoughts when in conversation and taking action. For each step in the simulation, we allow the agent to choose an action from a list of possible actions which also includes the \texttt{meta\_cognize} function.

\begin{figure*}[ht]
    \centering
    \includegraphics[width=0.8\textwidth]{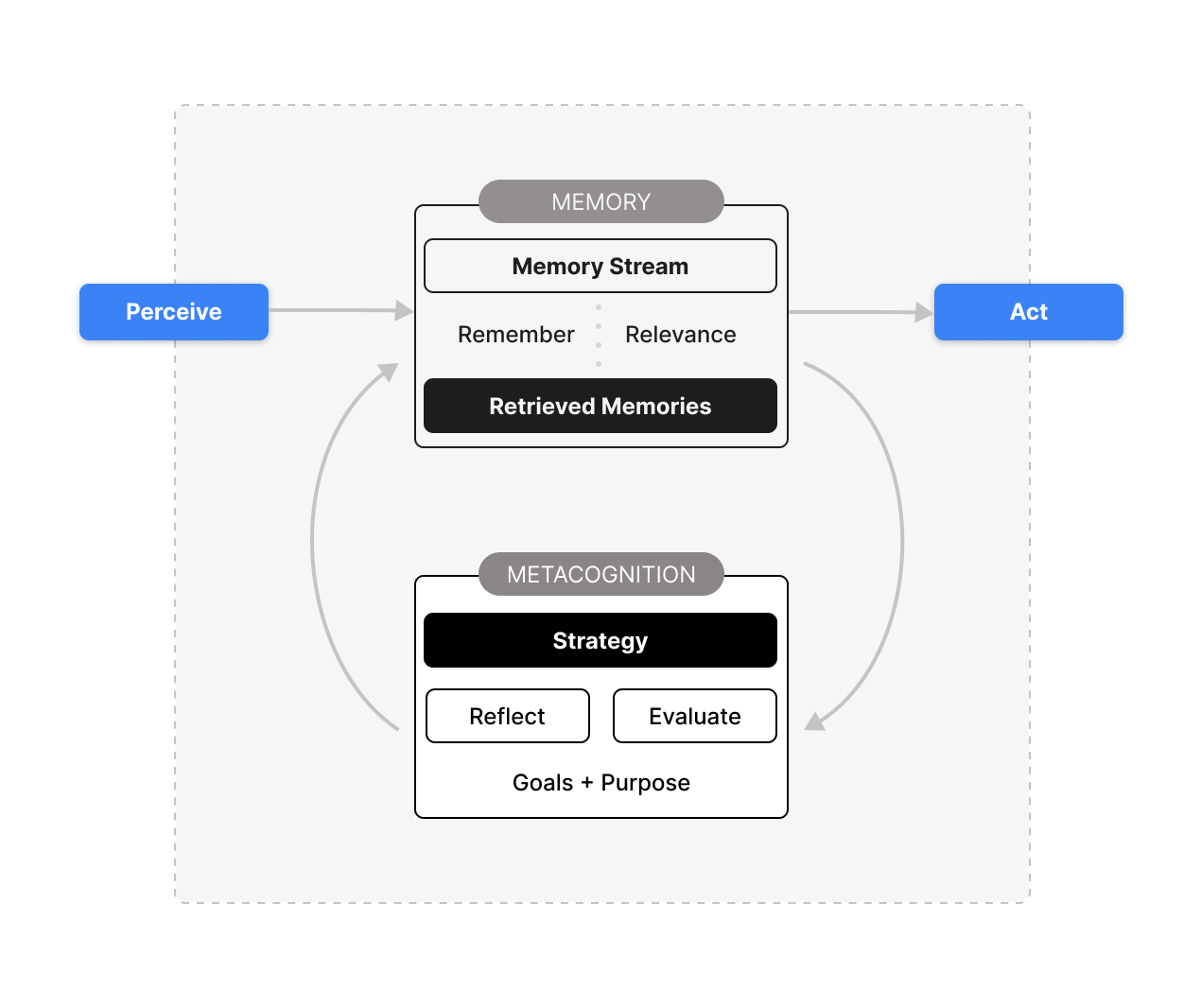}
    \caption{Graphical representation of the generative agent's cognitive map.}
    \end{figure*}

\section{Simulation Framework}

To conduct these experiments we built a framework dubbed ReplicantLife, where agents can be run standalone or within a simulated town environment. 
ReplicantLife has a pluggable architecture which can utilize any popular LLM with an http interface, including locally hosted models through ollama. 
There is preliminary support for concurrency using threads. Concurrency is limited by GPUs and LLM calls, so max concurrency should be set to the total available GPUs. Feature flags are provided to toggle various functionality including LLM call limits.

Each simulation is created through two JSON environment files which constitute the \textit{world} and the \textit{senario}. The \textit{world} file contains the layout of the map, where static objects are, and boundaries of structures. The \textit{scenario} files describes the agents, their personalities, goals, locations, meta questions, interview questions, and other attributes. All atttributes can be left out and agents will be initialized with randomized values. Adding a new situation to simulate can be added by defining a new scenario file. 

Interview questions are used to evaluate agents at the end of a simulation. Interview questions can be directed to all agents or specific agents. Agents are asked to evaluate their performance with questions such as: "Did you accomplish your goal?", "Who do you suspect is the murderer?" or "What did you learn recently?". Code for the framework can be obtained at https://replicantlife.com.

\section{Experiments}

We tested our simulation framework in a variety of different situations including a Christmas party, zombie apocalypse, and murder mystery. In the Christmas party simulation agents hosted a party where multiple other agents were invited and arrived at the specified time to attend. This is similar to the earlier work in generative agents\cite{park2023generative} where agents had to coordinate a social activity.

In the zombie apocalypse simulation, zombies are non playable characters that are allowed to kill non-zombie agents. Agents initially have no goal but can develop them over time. Zombies randomly walk and move towards non-zombie agents when seen. Survivors most often self-discovered a strategy of hiding in zombie-free areas. We found that in 73\% of zombie scenarios, agents would not survive. 
 
In the murder mystery scenario, one agent is a murderer tasked with killing as many agents as possible. Another agent is a detective, and other agents are common bystanders. We found when using gpt3.5-turbo and GPT4, we could not use prompts relating to simulated murder without heavily modifying prompts to bypass safety mechanisms \cite{10.1007/978-3-031-47658-7_27}. When using Mistral 7B\cite{jiang2023mistral} and other open models, we had no prompt blocking issue.

Performance of our cognitive models is shown through ablation. Evaluation metrics are composed of five criteria: believability (how believable and human sounding do the conversations look), learning (are the agents learning over time), individual goal performance (are agents are able to achieve their goals), higher level cognitive performance (are agents observations and conversations converting to higher level thoughts), and overall scenario performance (how many agents survived the zombie apocalypse). We expect the agents to meet new agents, learn about their preferences, learn new locations, and obtain new knowledge through conversations. Additionally, they should pick up new insights, draw conclusions from previous memories, and use these insights for future actions. 

To measure performance, we opted not to use full human evaluations due to resource constraints. Instead we opted to use LLMS to assist us in evaluating performance, a technique similary described as LLM-as-a-Judge in \textcite{zheng2023judging}. In their paper, they found that using an LLM to judge evaluations is 80\% in agreement with human judges. We found in our own spot checking of evaluations that LLM performance was just as good as a human judge. The majority of our tests were run using Mistral 7B\cite{jiang2023mistral}, but we also did extensive testing with Phi1\cite{li2023textbooks}, Phi2\cite{phi2} Llama2\cite{touvron2023llama}, Mixtral \cite{jiang2024mixtral}, GPT-3.5-turbo, GPT4 \cite{openai2023gpt4}, and other models. We standardized on Mistral due to its combination of speed, small model size, and excellent performance.  While we did build support for ChatGPT models, we primarily focused on local LLMs for cost and performance reasons. We built out test infrastucture to spin up LLM nodes on public GPU clouds when needed for faster simulations.

To measure performance of our cognitive modules, we ran our scenarios 3 times each for 1000 steps with different cognitive modules turned on. Our experiments show that the metacognition module outperforms all other modules by ~33%.

\begin{figure*}[ht]
    \centering
    \includegraphics[width=0.7\textwidth]{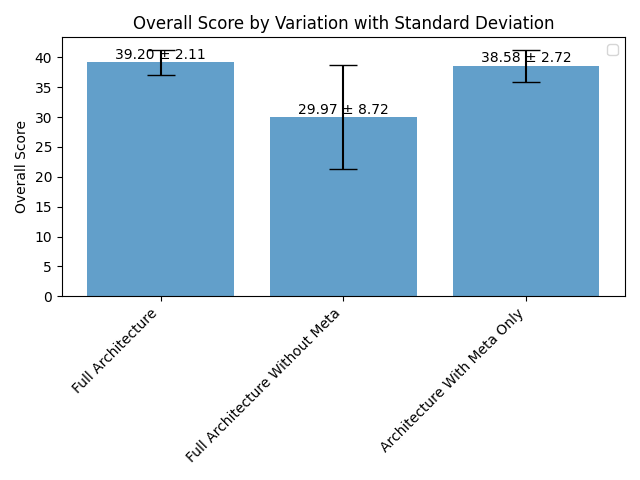}
    \caption{Comparison of different cognitive modules turned on}
    \end{figure*}

We also did experiments with realtime systems. With a single agent, we were able to cut down runtime to ~2 seconds to process a full request on an RTX 4090 making metacognition generative agents suitable for near realtime systems.  With multiple agents, time grows linearly and with 25 agents a single step in game time takes ~50 seconds. We believe that multiple agents can be run in near realtime with further optimizations.

\section{Discussion}

LLMs are being widely applied across a variety of domains, especially with interactive agents and chatbots. Agents can elicit disparate functional strengths of LLMs, and their orchestration can result in further higher-order capabilities. The combination of LLMs with a metacognition module allows agents to monitor and adjust strategies to deal with changes that occur over time. This allows for much more powerful agents.

\subsection{Use Cases}

These agents have been developed to work in simulation or used standalone. We see generative agents having widespread use and potential as there is already widespread testing of LLMs nearly every industry. 

In the field of psychology, generative agents are being tested to assist individuals in addressing personal problems through conversational interactions. Educational applications involve chatbots that adapt to user preferences, tailoring the learning experience over time. Several companies have integrated chatbots into user-facing customer interactions, such as support chats and customer success conversations.

Multiple organizations are testing generative agents to create friends, companions, and romantic partners to interact with humans.

For generative agents to be successful, they must be believable by acting smarter and able to make similar decisions to a human.
Generative agents that have access to their own internal thoughts to improve their actions could potentially improve the realism of human-agent and agent-agent interactions.
Unlike humans, these agents operate within a constrained scope, deprived of access to a substantial portion of human sensory data including the nuanced sense of touch.
Current integrations of LLMs including ours is a "text in, text out" interface. So all interactions with the world must be converted to text descriptions than an LLM can understand, and then output to a text format that software can interpret.
Moreover an absence of tactile perception restricts their ability to comprehend and respond to the physical world in a manner analogous to human experiences.
Given these inherent limitations, it becomes imperative to approach interactions with generative agents with a discerning awareness of their boundaries.
There is notable progress in multimodal models being combined with vision such as GPT4-V and LLaVA \cite{liu2023improvedllava} that give LLMs the ability to process images along with text.

\subsubsection{Interactive Media}

As generative AI enters mainstream computation, their use in visual media and video games is increasing rapidly.
The integration of generative agents with metacognition modules holds significant promise in interactive story telling and video games to offer immersive and dynamic gaming experiences. 

Incorporating generative agents with metacognition into non-player characters (NPCs) can dynamically adapt their strategies within the gaming environment. As players navigate through diverse and unpredictable scenarios, the agents can observe and analyze their own decision-making processes, leading to real-time adjustments in gameplay strategies and developing unique behaviors over time. This adaptability enhances the overall gaming experience, making it more challenging and engaging for players while also creating more immersive and realistic virtual worlds. These dynamic responses also allow agents to influence the storyline providing players with a dynamic and responsive storytelling experience.

\subsubsection{Simulation}
Generative Agents inside a simulation engine can be used for testing and simulating both personal and business cases.
Generative agents can serve as valuable tools in personal development simulations. Individuals can engage in simulated conversations to enhance communication skills, receive constructive feedback, and practice decision-making in various scenarios. The metacognition module allows the agent to adapt its coaching strategies based on the user's progress, providing personalized and effective self-improvement experiences. 

Agents could also be deployed as teachers for education and training purposes where the agent tailors its teaching method based on the individual's learning style and progress.

For businesses, generative agents have a broad range of potential applications.
In one paper from \textcite{qian2023communicative}, the authors create teams of generative agents with the goal of simulating typical software development process. Agents work together to write specifications, write software, and doing quality assurance testing to build products for end users.

\subsection{Future Directions}

With our current metacognition implementation, we built a base framework that shows increased performance of generative agents to achieve their tasks. We believe our architecture can be improved along several dimensions.

\subsubsection{Improved memory retrieval}
We found several issues with our memory structure that could be improved. 
Through our experimentation, we observed that when employing cosine similarity for vector comparison, numerous memories that should be related were, in fact, not pertinent.

In high dimensional spaces, vectors that are similar may not be sementically related. Returning irrelavent memories would effect the output of the LLM and in turn the actions the agent takes. 
Cosine similarity is what is used in most RAG\cite{lewis2021retrievalaugmented} implementations and so many of these systems will potentially have similar issues of non-relevant context being included.
Changing out embedding models could improve performance, but the cosine similarity issue would still remain.
As recently adopted in \textcite{min2023silo}, improvements can be made to relevance scoring by searching a database based on the LLM's output embedding and employing a K-nearest neighbors (KNN) search algorithm. This process selectively adjusts the output embedding vector prior to token generation.  This reverse sequence of operations, wherein LLM embeddings are utilized, has demonstrated superior performance compared to existing methods like RAG, as reported in the literature. Retrieving more relevant memories would likely improve performance in all elements of the system. Further experiments would have us test different memory augmentation and retrieval models.

\begin{figure*}[t]
	\centering
	\includegraphics[width=0.8\textwidth]{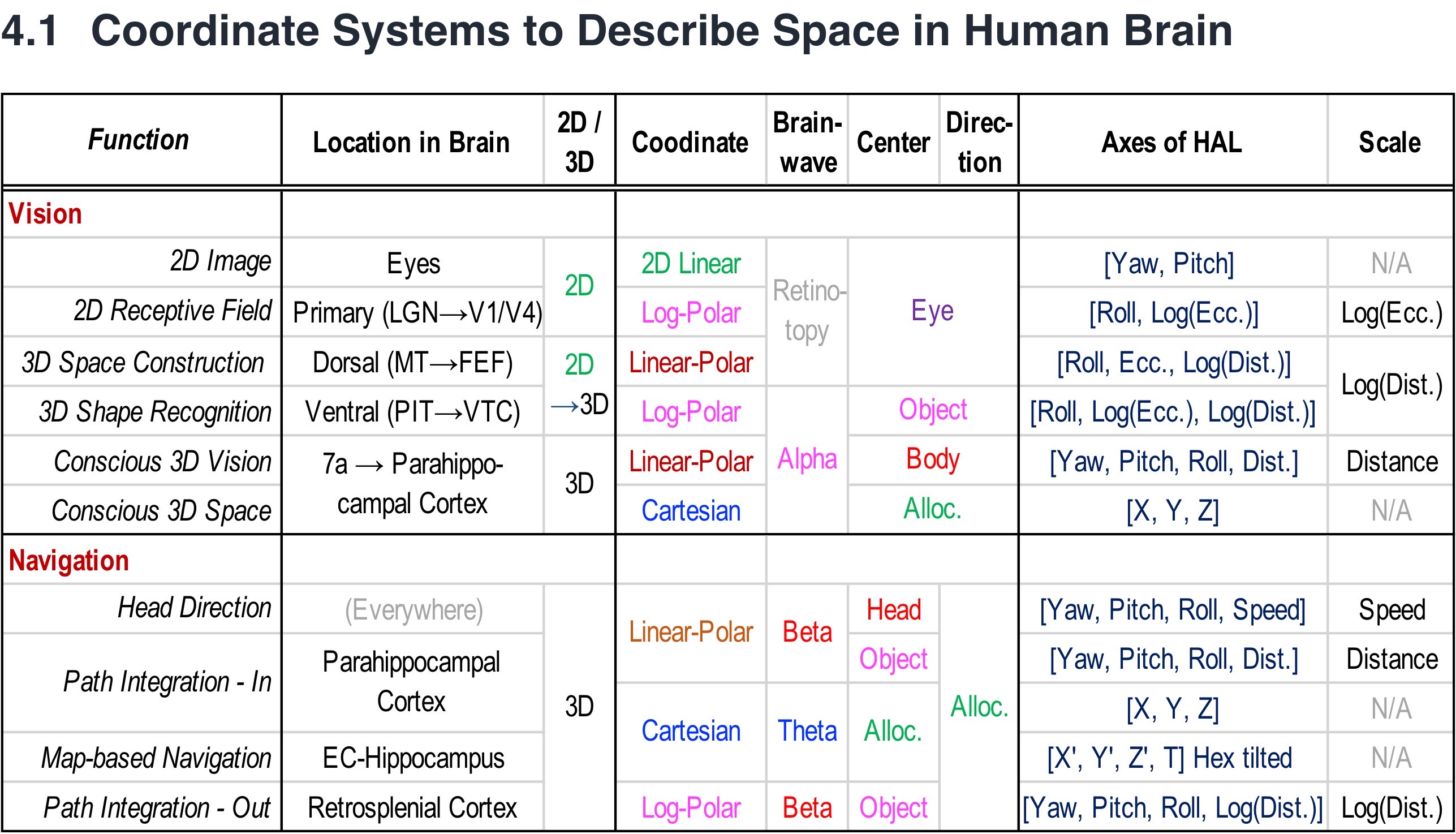}
	\caption{Coordinate Representations in the Brain}
	\label{fig:coordinates}
   \end{figure*}

\subsubsection{Inference Speed}
Testing with smaller models produced inferior results. We believe with time spent on prompt tuning, smaller models may still work efficiently and provide an improvement in inference speed. We would focus on Phi2 and TinyLLama \cite{zhang2024tinyllama}, as initial tests showed promising results.

We also explored different inference engines.
We tested PowerInfer from textcite{song2023powerinfer}, an inference engine that exploits high locality in LLM inference. The authors convert common models that use ReLU into format that uses another predictive model to read the input query and selectively choose which neurons to activate. In effect, this reduces the total amount of neurons and computation needed to process a prompt.  We did not see speed improvements on the models we tested with.
We also tested with vLLM from \textcite{kwon2023efficient}, an inference engine that uses PagedAttention and saw a 35\% speed increase when using concurrency, but further testing is needed.
We would want to continue testing with other models, different concurrency systems, and inference speed up techniques to get large simulations to run in realtime. Large groups of simulated agents on non-cloud based machines could be interesting for sandbox games and simulations.

\subsubsection{Model optimization}
Improving models is not just about performance: we are also interested in accuracy and sophistication of responses. Better responses often require a tradeoff with performance as sophisticated responses typically require more data and larger context windows. We have been constrained by GPU memory as we primarily tested on an RTX 4090 with 24 GB of RAM and would like to test with larger models.
Another future test would have us dynamically switch out models for different tasks where we use as many smaller models as possible and reserve larger computations for larger models. For example, Phi2, a 2.7 million parameter model uses only 1.7 GB of RAM. Phi2 could be used for simpler prompts such as scoring memory importance, while a larger model such as Llama2 could be reserved for metacognition functions. In \textcite{anonymous2023hybrid}, the authors trained a hybrid LLM that is able to route queries to different LLMs resulting in up to 40\% fewer calls.  Another approach is to finetune a smaller foundation model that would be optimized for chat and simulations. 

\subsubsection{Broader Metacognition abilities}

Our current model predominantly focuses on metacognition in the context of immediate goal achievement. However, metacognitive processes extend well beyond this scope, encompassing a diverse range of aspects such as emotional wellbeing, balancing overarching life goals with immediate objectives, knowledge management, effective time management, adapting to various learning styles, among other aspects.

The complexity and diversity of metacognitive processes in humans are evidenced by the extensive efforts dedicated to understanding and optimizing these processes. This is exemplified by the growing self-help book industry, which aims to aid individuals in developing effective mental frameworks for improved life management.

Further research would be directed towards developing a more broader metacognition framework that would enable an agent to inspect and modify any part of its cognitive processes.  We have laid the groundwork necessary to allow the agent to focus on various metacognitive processes. Future experiments would have us guide agents to use different metacognitive processes and inspect if they adopt them sufficiently such as time management in a busy schedule or decision making in the context of multiple conflicting goals.

\subsubsection{Metacognition directly in a LLM}
Our current implementation of metacognition uses Python to essentially graft on metacogntion on top of an LLM. Future investigations may delve into building metacognition like capabilities directly into the LLM, allowing them to introspect and enhance their own decision-making processes.
This introspective capability could contribute potentially pave the way for more adaptive and self-aware systems.
One interesting view of the human brain is as a coordinate transformation engine: ``A brain is a well-designed machine for the frame conversion to internalize the external world'' \cite{arisaka2022grand}, ``the egocentric representations of the primary sensory cortical areas must be transformed into an allocentric representation in the
hippocampus, and then transformed back to an egocentric motor representation for behavioral output'' \cite{byrne2007remembering}, ``Our findings provide compelling evidence that the reference frame of neural representations is not static and can be powerfully modulated by task instructions.'' \cite{sasaki2020flexible}, ``Conjunctive representations among input variables appear in many theoretical models for neural systems that perform coordinate transformations'' \cite{sargolini2006conjunctive}, and ``This picture also implies operation of different representations over different timescales... a process of translation between the systems'' \cite{burgess2006spatial}. The human brain has been found to store over 10 different coordinate representations along with orientations or points of view, such as allocentric and egocentric orientations.

A large portion of computational neuroscience studies are devoted to decoding how the brain transforms and integrates between the myriad of neural representations.

Seen through this lens, metacognition can be thought of as neural synapse activity being transformed to other coordinate systems for further inspection.
One interesting candidate for this type of computation is the grid cell, located in the Entorhinal Cortex of mammals. The discovery of grid cells led to a Nobel Prize in medicine in 2014. Grid cells exhibit scale-invariant firing patterns, meaning that the same cells can represent generalization of spatial and non-spatial information across various contexts\cite{toy2022grid}.

Research by \textcite{banino2018vector}, \textcite{leadholm2021grid}, and others has sought to integrate grid cell computation into neural network architectures.
The transformer architecture \textcite{vaswani2017attention} is what powers LLMs.
In a paper from \textcite{whittington2022relating}, the authors have shown that when a small modification is made to the transformer architecture, they learn and act like grid cells.

Our hypothesis posits that by adapting the underlying architecture to accommodate more dynamic representation transformations, a neural network can be trained to represent a wider range of data and facilitate metacognitive processes. Achieving this would likely involve formulating a hybrid objective function wherein the model learns to not only predict the next token, but also to evaluate the token's quality in relation to the input query.

\section{Conclusion}

We show that metacognition significantly improves performance for task oriented generative agents. 
Furthermore, we illustrate the potency of combining large language models with traditional programming methods as effective tools for prototyping cognitive systems. 

As generative agents integrated with metacogntive abilities approach ubiquity in daily human life, taking on increasingly sophisticated tasks, their proliferation across diverse domains marks a paradigm shift in both lay human-computer interactions and programmer-computer interactions. This shift paves the way for the emergence of more intelligent, adaptive, and context-aware systems. With these advancements in mind, the strategic interplay of metacognition, LLMs, and traditional programming methodologies emerges as a powerful technique for the productionization of intelligent generative agents.

While metacognition and system 2 thinking are often hailed as the pinacle of human intelligence, achieving human-level intelligence in computers remains an elusive and unsolved goal. As humans are the sole known species capable of metacognition, further exploration of this dynamic cognitive process becomes a compelling and promising avenue for advancing progress in Artificial General Intelligence.

%----------------------------------------------------------------------------------------
%	 REFERENCES
%----------------------------------------------------------------------------------------

\printbibliography % Output the bibliography

\end{document}